# 3D photoionization models of nova V723 Cas

L. Takeda,[1⋆] M. Diaz,[1] R. Campbell[2] and J. Lyke[2]
[1]IAG, Universidade de São Paulo, Rua do Matão, 1226 , Sao Paulo, SP 05508-900, Brazil
[2]W. M. Keck Observatory, 65-1120 Mamalahoa Hwy., Kamuela, HI 96743, USA

Accepted 2017 August 31. Received 2017 August 30; in original form 2017 June 9.

## ABSTRACT

We present modeling and analysis of the ejecta of nova V723 Cas based on spatially resolved IR spectroscopic data from Keck-OSIRIS, with LGSAO (adaptive optics module). The 3D photoionization models include the shell geometry taken from the observations and an anisotropic radiation field, composed by a spherical central source and an accretion disk. Our simulations indicate revised abundances $log(N_{Al}/N_H) = -5.4$, $log(N_{Ca}/N_H) = -6.4$ and $log(N_{Si}/N_H) = -4.7$ in the shell. The total ejected mass was found as $M_{shell} = 1.1 \times 10^{-5}\ M_\odot$ and the central source temperature and luminosity are $T = 280,000\ K$ and $L = 10^{38}\ erg/s$. The 3D models are compared to basic 1D simulations to demonstrate the importance of using more realistic treatments, stressing the differences in the shell mass, abundances and characterization of the central source. The possibility of V723 Cas being a neon nova and the puzzling central source features found are discussed.

**Key words:** novae, cataclysmic variables – stars: individual(V723 Cas) – accretion, accretion discs – stars: abundances – line: formation – (stars:) white dwarfs

## 1 INTRODUCTION

Novae are cataclysmic variables that present one or more eruptions driven by thermonuclear runaways on the surface of a white dwarf. These eruptions occur due to the mass transfer of hydrogen-rich gas from the secondary star to the white dwarf (primary star). Nova eruptions eject envelopes of typically $10^{-6}$ to $10^{-4}\ M_\odot$ of gas, composed by matter of both primary and secondary stars (Warner 1995). Each nova presents different shell properties, specially in terms of geometry, density condensations and velocity field. Despite that, the current photoionization models of nova shells usually assume spherical geometries and power-law density functions, or, in a few cases, bipolar geometry and random gaussian mass distribution (Schwarz et al. 2007). The discrepancies between the models and the real structures can be noticed by the results from different models for the same objects. For instance, the ejected mass and heavy metal abundances estimated for novae by different authors are often inconsistent, and lead to distinct interpretations of the physical properties of the nova (see Hayward et al. 1996; Austin et al. 1996; Arkhipova et al. 1997; Vanlandingham et al. 2005, as an example for V1974 Cyg). One solution for this problem is to map the actual mass distributions of nova shells through spatially resolved spectroscopy and use it in 3D photoionization models.

The spatially resolved spectroscopy can provide the projected emission from the actual 3D distribution of the envelope gas. Unfortunately, the nova shells have small angular diameter and low surface brightness so they are difficult to observe. When the shell has already expanded, it can achieve only a few arcsec, because the expansion lowers the density and moves the gas away from the ionizing source, making its emission progressively fainter. Another observational difficulty is the angular resolution and dynamical range needed to detach the central source emission from the envelope emission. Therefore, the observations of nova shells are usually made from space, or made with the help of adaptive optics modules. These spectroscopic data can be transformed into a 3D flux distribution map, assuming a constant expansion velocity. As the flux of permitted transition lines is proportional to the squared density (Osterbrock & Ferland 2006), the flux distribution can be converted to a mass distribution map. This 3D mapped shell can be used in 3D photoionization codes to estimate more accurate values of temperature and luminosity of the central source, total ejected mass, chemical abundances, among other properties of the nova system, and, consequently, help to elucidate unresolved problems concerning novae. The derivation of accurate values of ejected mass for classical and recurrent novae, for instance, is key to determine if the white dwarf in novae could be a progenitor of type-Ia supernovae. More accurate estimates of chemical abundances are also fundamental to explain the puzzling over-abundance of iron-peak elements in some novae.

⋆ E-mail: larissa.takeda@usp.br

In this paper we present a global approach to the diagnostic of nova V723 Cas, using 3D photoionization models based on integral field spectrograph (IFS) data and several multiwavelength spectra. V723 Cas is a classical nova (with only one eruption registered) and was detected on August 24th, 1995 (Hirosawa et al. 1995). It is an unusually slow nova, with $t_{3,V}$ = 173 days and $t_{3,B}$ = 189 days (Ohshima et al. 1995) and it presented the longest nuclear burning activity registered for novae. It also presented an extremely hot central source (Ness et al. 2015) for a slow nova, as detected by X-rays observations and presented overabundance of neon, detected using visible spectra (Iijima 2006). Along with these peculiarities, V723 Cas also presented spectral features consistent with a complex geometry shell that were later confirmed by spatially resolved spectroscopy in IR (Lyke & Campbell 2009). These observations were performed with Keck-OSIRIS near-infrared integral field spectrograph and an adaptive optics module from 2005 to 2008. Our models are mainly based on the 2008 data, when the nova was in the nebular stage, with spectral coverage of 1955 – 2055 $nm$, 2121 – 2229 $nm$ and 2292 – 2408 $nm$. V723 Cas was intensively observed with other different instruments and in different wavelengths during its eruption. We gathered the available data in IR, X-rays, optical and radio to constrain our models and compare our simulations to all observables at hand.

## 2 RAINY3D PHOTOIONIZATION MODELS

We used an updated version of RAINY3D code (Moraes & Diaz 2011), which is a photoionization code that models any 3D geometry of gas distribution and any shape of ionizing radiation field. It runs version 13.04 of Cloudy (Ferland et al. 2013) as a subroutine and compares the simulated emissivities to the 2D observational data, providing the best fit parameters. Cloudy is responsible for the calculation of matter-radiation interaction, including thermal and statistical equilibria, ionization and neutralization balances, heating and cooling processes and NLTE populations. RAINY3D creates the geometry and the mass distribution in 3 dimensions and organize the input parameters to run Cloudy in an adaptive mesh. Then, RAINY3D integrates and projects the shell emission and compares them to the observed distribution through a relative $\chi^2$ method.

We have performed a series of photoionization models with different degrees of complexity. The first models were characterized by spherical shell geometry with power-law density function and uniform ionizing radiation field (standard 1D models). The second grid of models had 3D geometry based on IFS observations and uniform ionizing radiation field. The last models included both 3D geometry based on observations and the contribution from a disk-shaped ionizing source. All models had the input parameters listed in table 1, of which only 3 were initially free parameters (temperature and luminosity of the central source and the shell hydrogen mass). The input was based on multiple 2D observations in the IR, optical and X-rays, that constrained the properties of the central source and the shell. The details of our models are described in the next subsections.

### 2.1 Observed fluxes

Our models are mainly based on 2008 integral field spectroscopy data from Keck-OSIRIS and *Swift* X-ray data taken in the same year (Ness et al. 2008). In order to flux calibrate the 2005 OSIRIS spectra, we used the 2003 2MASS magnitude in K band and synthetic photometry derived from our spectra using the standard K filter transmission curve (Cutri et al. 2003). No significant broadband decay from 2003 to 2005 was considered in the K-band. We then calibrated the 2008 OSIRIS data using the 2005 calibration above and the [Si VI] relative fluxes from Lyke & Campbell (2009). We have estimated an uncertainty of ∼ 40% for the flux calibration by extrapolating the fit of the flux decay from 2005 to 2008. We have adopted a distance to V723 Cas $d = 3.85$ $kpc$ (Lyke & Campbell 2009) and $E(B-V) = 0.5$, as an average of different values in literature, to obtain the intrinsic emitted fluxes. We considered the values of distance and $E(B-V)$ found by other authors with different techniques to estimate the error in our fluxes. Distance values of 4.2/4.0 $kpc$ (Evans et al. 2003), 2.80 $kpc$ (Iijima 2006) and 2.70 $kpc$ (Ness et al. 2008) were considered. For $E(B-V)$, we considered $E(B-V) = 0.60$ (Gonzalez-Riestra et al. 1996), $E(B-V) = 0.45$ (Munari et al. 1996), $E(B-V) = 0.54/0.57$ (Chochol & Pribulla 1997), $E(B-V) = 0.20/0.25$ (Rudy et al. 2002), $E(B-V) = 0.78/0.60$ (Evans et al. 2003), $E(B-V) = 0.57$ (Iijima 2006) and $E(B-V) = 0.35$ (Hachisu & Kato 2014). We used the extreme cases of largest distance and highest $E(B-V)$ and smallest distance and lowest $E(B-V)$ to set the limits to intrinsic emissivity values. The errors in flux calibration lie within this range.

### 2.2 Input abundances

Iijima (2006) derived the abundances of N, O, Ne, Ar and He from the optical spectra of V723 Cas. He used the fluxes of He I and He II, [N II], [O III], [Ne III], [Ne IV] and [Ne V] and [Ar III], [Ar IV] and [Ar V] to make the estimates. Evans et al. (2003) derived the relative abundances of Al/He, Ca/He, S/Si, Al/Si and Ca/Si from the IR spectra of V723 Cas, using the fluxes of [Al VI], [Al V], [Ca IV] and [Ca VIII] lines. We used Iijima's estimation for He abundance to calculate Al/H and Ca/H from Evans's data. For the Si/H ratio, we used an average of the values obtained through Ca and Al abundances. For the other elements, we adopted average values for novae (Gehrz et al. 1998). Although these abundances are based on simplistic models, we used them as an initial guess for our models, aiming to constrain the physical properties of the central source and shell. In a second step, the Al, Ca and Si abundances were fine tunned as needed to match both the integrated fluxes and observed structures.

### 2.3 The anisotropy of the ionizing radiation field

The OSIRIS data showed distinct shell morphologies for different emission lines. Overall, the shell is formed by two main structures: an equatorial torus and polar nodules. For the [Si VI] 1.963$\mu m$ and [Ca VIII] 2.321$\mu m$ bands, the torus is more prominent, and for the [Al IX] 2.044$\mu m$, the polar emission is stronger. The ionizing potential for these levels are, respectively, 205.27 eV, 147.24 eV and 330.21 eV. This effect, as already pointed out by Lyke & Campbell (2009), reveals an anisotropy in the ionizing radiation field, with higher ionization in the poles and less in the equator, and it has been detected in other novae, as HR Del (Moraes & Diaz 2009), DQ Her (Petitjean et al. 1990) and FH Ser (Gill & O'Brien 2000).

**Table 1.** RAINY3D input parameters for V723 Cas models

| Parameter | Value | Reference |
|---|---|---|
| distance | 3.85 $kpc$ | Lyke & Campbell (2009) |
| E(B-V) | 0.5 | see subsection 2.1 |
| $T_{CS}$ | $2.8\times10^5 - 3.8\times10^5$ $K$ | Ness et al. (2008) |
| $L_{CS}$ | $5\times10^{36} - 2\times10^{38}$ $erg/s$ | Ness et al. (2008) |
| $r_{in}$ | $1.03\times10^{16}$ $cm$ | this paper |
| $r_{out}$ | $2.50\times10^{16}$ $cm$ | this paper |
| $\alpha_{p.c.}$ [1] | 20° | this paper |
| $M_{env}$ | $5\times10^{-6} - 1\times10^{-4}$ $M_\odot$ | Iijima (2006),Heywood et al. (2005) |
| $\rho_{p.c.}$ [2] | $10^3$ $cm^{-3}$ | this paper |
| $\rho_{ring}$ [2] | $10^{3.8}$ $cm^{-3}$ | this paper |
| $\rho_{clumps}$ [2] | $10^5$ $cm^{-3}$ | this paper |
| $\rho_{diffuse}$ [2] | $10^{3.5}$ $cm^{-3}$ | this paper |
| $log(N_N/N_H)$ | -2.8 | Iijima (2006) |
| $log(N_O/N_H)$ | -2.2 | Iijima (2006) |
| $log(N_{Ne}/N_H)$ | -2.5 | Iijima (2006) |
| $log(N_{Ar}/N_H)$ | -4.3 | Iijima (2006) |
| $log(N_{He}/N_H)$ | -0.74 | Iijima (2006) |
| $log(N_{Al}/N_H)$ | -6.98 | Evans et al. (2003) |
| $log(N_{Ca}/N_H)$ | -7.86 | Evans et al. (2003) |
| $log(N_{Si}/N_H)$ | -7.40 | Evans et al. (2003) |

[1] The $\alpha_{p.c.}$ value corresponds to the aperture angle of the polar caps.
[2] The $\rho$ values are the maximum hydrogen density adopted for the structures.

We speculate that, as for the case of HR Del, the V723 Cas anisotropic field may be a consequence of the restoration of the accretion disk as the pseudo-photosphere contracts. We estimated the radius of the photosphere in $r = 1.34\times10^9$ $cm$, using the luminosity and temperature suggested by X-ray observations at the same epoch (2008). Comparing to the primary Roche Lobe radius, estimated in $r_{RL(1)} \sim 6.0\times10^{10}$ $cm$, we can conclude that the disk radius, that is typically 70% of $r_{RL(1)}$, is greater than the photosphere radius at that time.

We have included a tool in RAINY3D to calculate models with a disk-shaped ionizing source, but it can be adapted to any other shape. Instead of considering an ionizing flux independent of the polar angle ($\theta$), we adopted $F'(\theta) = F\cos(\theta)$. The integrated flux in solid angle in both sides of the disk is equal to the total luminosity of the central source. This disk-shaped source can also be added to the spherical central source, in the case that both are significant. We also included the option of adopting polar symmetry in order to reduce the computational time.

## 2.4 Central source atmosphere model

The X-ray data from Ness et al. (2008) indicate that the central source temperature is in the range of $2.8\times10^5 - 3.8\times10^5$ $K$ and the luminosity in the range of $5\times10^{36} - 2\times10^{38}$ $erg/s$. This is an extremely and unusually hot central source. A simple blackbody model was found to be an acceptable option for the high temperature central source continuum. We usually prefer the Rauch planetary nebulae nucleous models (Rauch 2003), that are more complex and realistic. But in this case, the central source temperature is higher than the limits of the Rauch models that consider metals up to Ni. The other options (pure hydrogen, hydrogen and helium) lacks the metals that strongly affects the ionizing continuum, providing higher fluxes than expected. Comparing the grids of Rauch models, it is evident that the presence of metals lowers the ionizing flux, and the blackbody model represent this feature better than the Rauch models without metals.

## 2.5 Shell geometry

As mentioned before, our models are based on the OSIRIS IFS data taken in 2008. With the central source emission subtracted from the data cubes, we assumed the expansion velocity was constant and transformed the wavelength axis in spatial axis. We assumed that the equatorial torus is circular, and that the elliptical shape is due to the system's inclination of $62^o$ (Lyke & Campbell 2009). Unfortunately, the shell recombination lines fluxes were too weak to map the density distribution. If the $Br\gamma$ was bright enough, we would estimate the hydrogen density along the cube. Since we could not directly measure the density, we combined the structures seen in the data cubes with the total shell mass estimated by Iijima (2006) and Heywood et al. (2005). A 3D shell composed by a low-density spherical shell, an equatorial torus with clumps and polar caps, was created as displayed in figure 1. We assumed that the ring and polar caps were denser than the diffuse spherical shell, and that the clumps were denser than the ring and caps. The insertion of clumps was necessary because the ring gas overlap due to the inclination was not enough to reproduce the ring luminosity contrast seen in [Ca VIII] and [Si VI] images. The luminosity asymmetry at the opposite sides of the ring observed in [Si VI] images also suggested the need for a denser component. The density contrasts were smoothed by a gaussian function in order to facilitate the convergence of Cloudy's calculations. The gas was distributed in a way that the total mass would be within the cited range. We are able to project the resulting 3D emissivity volumes to the line-of-sight, neglecting the line radiative transfer within the shell. This approximation certainly holds true for the forbidden lines analyzed here.

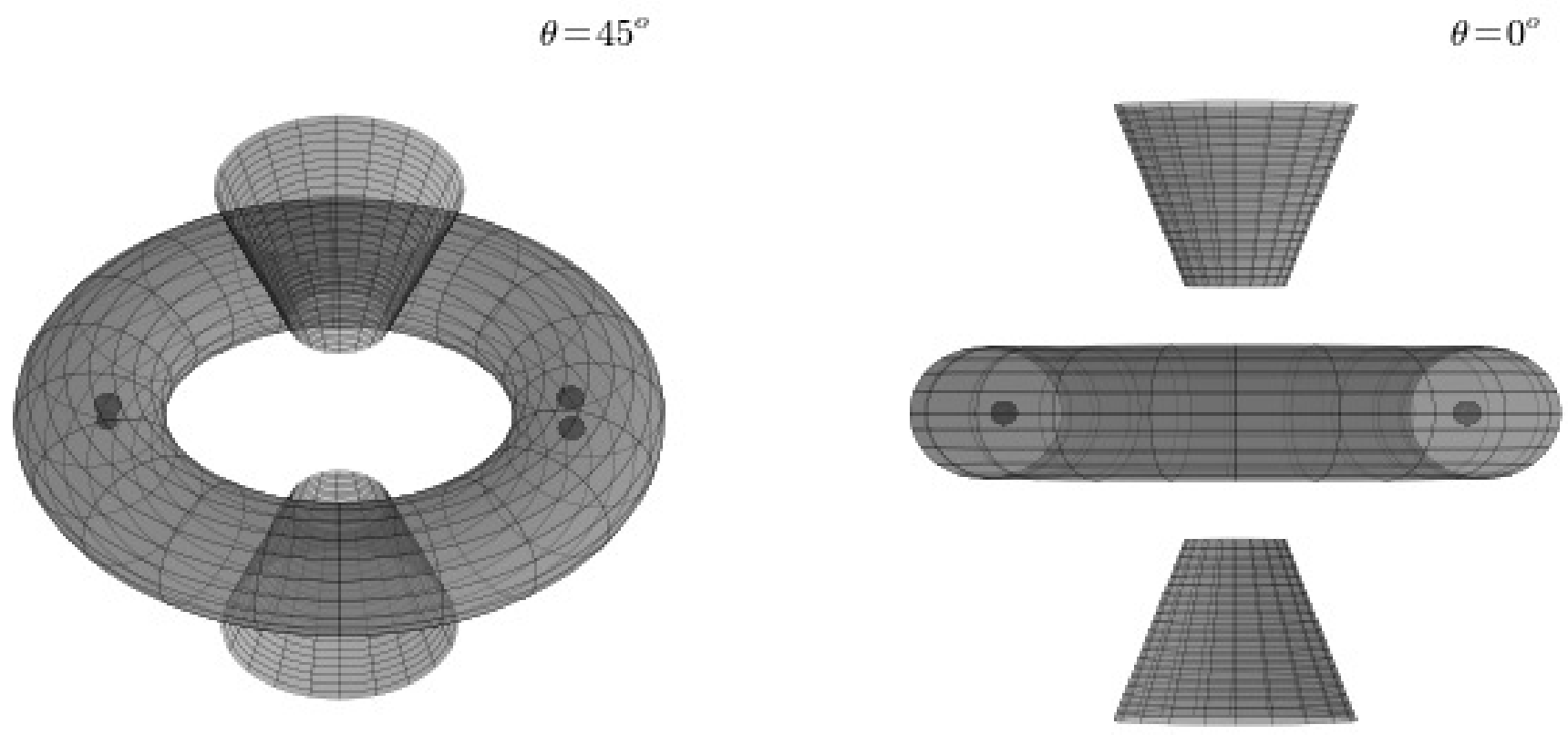

**Figure 1.** Input hydrogen gas envelope geometry, showed from different angles. Diffuse spherical shell was also added to the models, but not to the image to facilitate visualization. The dark spots show the position of the gas condensations

## 3 RESULTS

### 3.1 1D models

We started with 1D models, assuming power-law radial density profiles with $\alpha = 0.84$ and an uniform ionizing radiation field. We varied the luminosity of central source in $36.5 \leq log(L) \leq 37.5$ (L in $erg/s$), the temperature in $280,000 \leq T \leq 380,000$ ($K$) and the shell mass in $6.0 \times 10^{-5} \leq M_{shell} \leq 1.2 \times 10^{-4}$ ($M_{\odot}$). The results are displayed in figure 2, with integrated line fluxes as a function of temperature of the central source. The gray areas in the figure are the limits for the line fluxes considering the full range of distance and E(B-V) values and the columns are divided according to the central source integrated luminosity used for each model. For lower values of temperature and luminosity of the central source, the fluxes of [Al IX] and [Ca VIII] are in good agreement. But for the permitted transitions, the models show much higher fluxes than observed. The [Si VI] flux is too low for every model in this class. The best fit obtained for this configuration was for $log(L) = 36.5$ (L in $erg/s$), $M_{shell} = 6 \times 10^{-5}$ $M_{\odot}$ and $T_{CS} = 380,000$ $K$. When comparing the flux ratio (see tables in appendix) $F_{He}/F_{H}$ for all models a good agreement is found. This is expected, since both are recombination lines and, thus, the flux ratio is weakly dependent of shell temperature. For all models, $F_{Si}/F_{H}$ is at least $10^3$ times lower than the observed value, and for $log(L) = 37.5$ (L in $erg/s$), the $F_{Al}/F_{H}$ approaches the observed value. The $F_{Ca}/F_{H}$ ratios are lower than observed for all models, but are better fitted for $log(L) = 36.5$ (L in $erg/s$), unlike the case of the [Al IX]. $F_{Al}/F_{Si}$ and $F_{Ca}/F_{Si}$ are higher than observed, mainly because the Si fluxes are too low. To better fit the observed data, these models would need lower hydrogen mass and higher Si abundance. Thus, the results would point to the lower limit for the central source luminosity ($log(L) = 36.5$, where L is in $erg/s$). One other possibility is to largely increase the Si and Ca abundances and lower the Al abundance, so the models would better fit the observed data for higher luminosity values. However, this would imply in quite unusual abundance ratios.

### 3.2 3D models with uniform ionizing radiation field

Our second grid of models included the 3D geometry, but not the anisotropic ionizing radiation field. We assumed that the ring and polar caps were 100 times denser than the diffuse spherical shell, and that the clumps were 10 times denser than the ring and caps. We considered the ring and polar densities limits equal because the gas velocities are similar in both structures, indicating that the gas expansion and its density should be similar as well. We started with the lower limits of mass and central source luminosity. The results are displayed in figure 3 as integrated line fluxes as function of temperature and in table A4, as fluxes ratios. Again, the $F_{He}/F_{Br\gamma}$ ratio is close to the observed value, and both H and He fluxes are higher than the observed data and higher than the previous models. For all models, $F_{Si}/F_{Br\gamma}$, $F_{Al}/F_{Br\gamma}$ and $F_{Ca}/F_{Br\gamma}$ are lower than the observed values, because the $Br\gamma$ flux is 100 times higher than observed. Unlike the previous models, the $F_{Al}/F_{Si}$ ratio is in good agreement to the observed value, specially for $T = 300,000K$. But $F_{Ca}/F_{Si}$ is $\sim$ 10 times larger for the models, and the integrated flux of [Si VI] line is lower than observed.

Figures 4 and 5 show the emissivity profiles in polar and equatorial planes for H I, [Si VI], [Al IX] and [Ca VIII] lines. The structures formed by H I, Si and Ca lines are similar, presenting both ring and polar caps in same intensity. The structures in Al line are weaker, and do not correspond to the observed features. Given the difference in Al ionization potential we interpreted such a mismatch as the uncounted presence of an anisotropic ionizing field. It is interesting to notice that the clumps are more luminous than the torus for H, Si and Ca line, but it is less luminous for the Al line, due to the combination of local density and ionization. For the H, Si and Ca lines, the radial emissivity profiles follow the hydrogen density, presenting higher emissivity at the center of the torus, where the density is higher. For the Al radial profile, however, the emissivity is higher at inner radius, and it decreases as the distance from the central source increases.

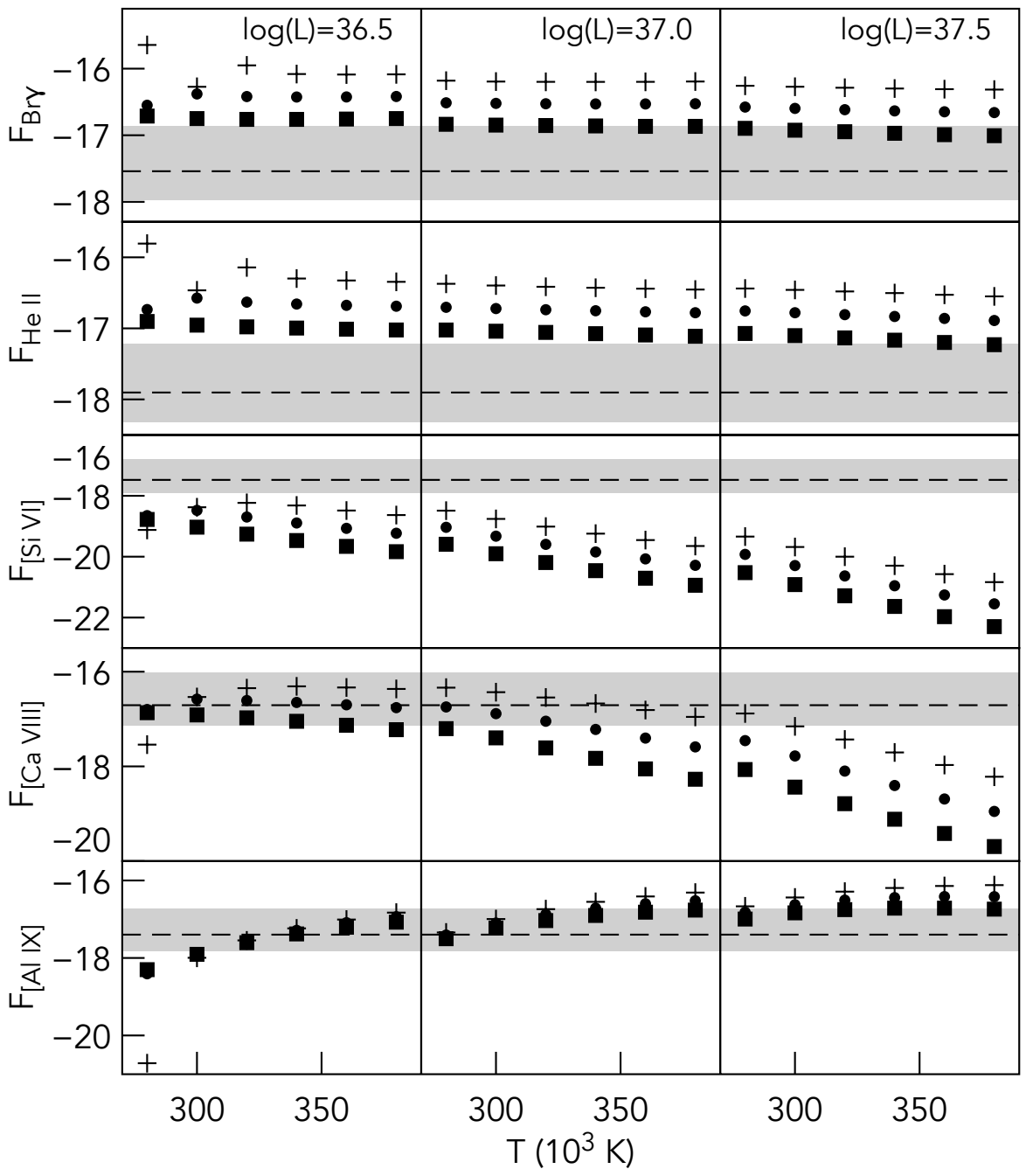

**Figure 2.** Total lines fluxes from RAINY3D models for a spherical shell geometry and uniform ionizing radiation field. The squares correspond to the models with $M_{shell} = 6.0 \times 10^{-5}$ $M_{\odot}$, the dots to $M_{shell} = 8.5 \times 10^{-5}$ $M_{\odot}$ and the crosses to $M_{shell} = 1.2 \times 10^{-4}$ $M_{\odot}$. The dashed lines correspond to the observed fluxes scaled to $d = 3.85$ $kpc$ and $E(B-V) = 0.5$, and the gray area correspond to the limits of the observed fluxes with different measurements of distance and E(B-V) values. The first column corresponds to $log(L) = 36.5$, the second to $log(L) = 37.0$ and the third to $log(L) = 37.5$, where L values are in $erg/s$. F is the log of observed flux in $erg/s/cm^2$.

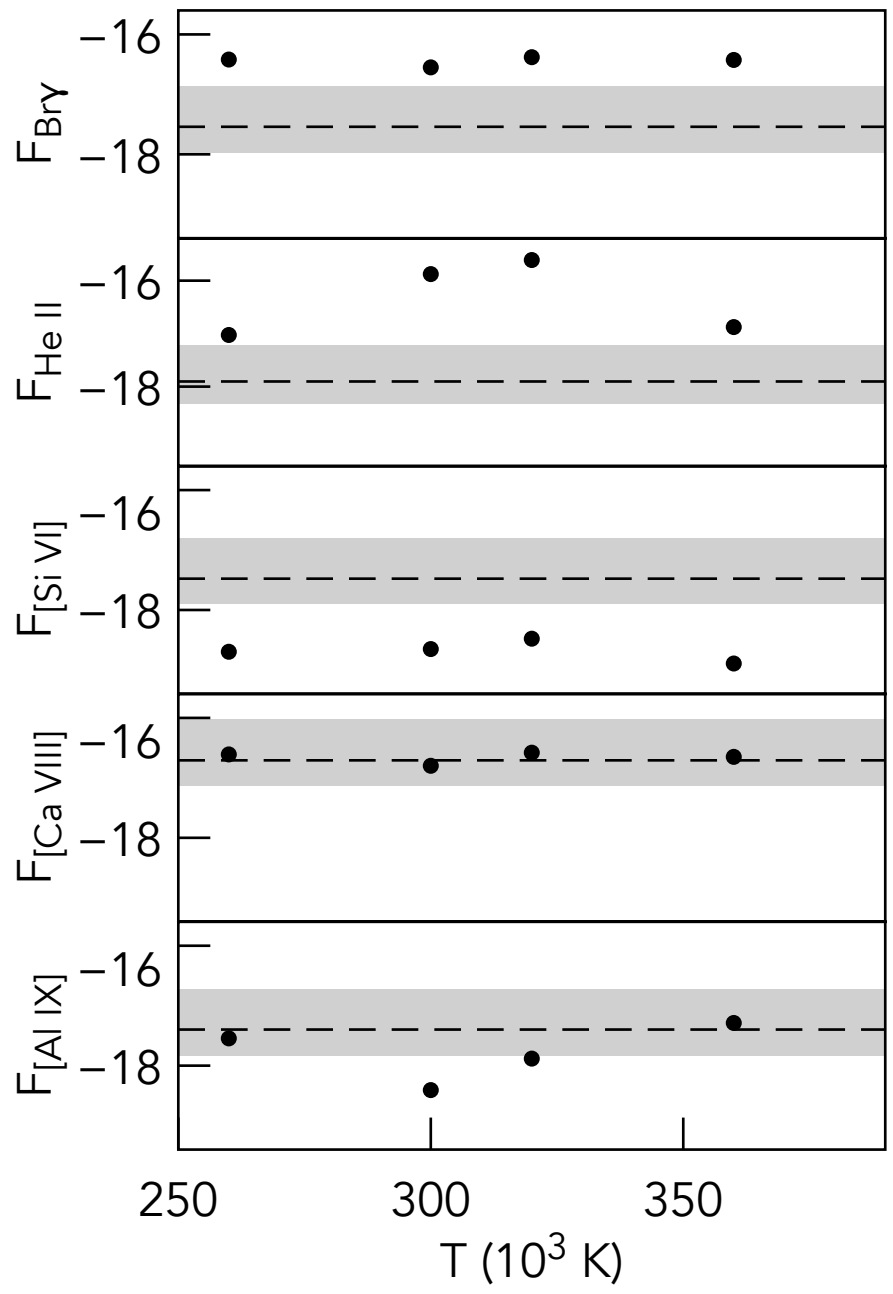

**Figure 3.** Same as fig. 2 for RAINY3D models with 3D geometry and an isotropic ionizing radiation field, with $M_{shell} = 6.0 \times 10^{-5} M_{\odot}$ and $L = 10^{36.5}$ $erg/s$. F units are log of observed flux in $erg/s/cm^2$.

### 3.3 3D models with anisotropic ionizing radiation field

Our third set of 3D models included both anisotropic ionizing radiation field and the observed geometry. The results indicate that the central source is hotter and more luminous than suggested by the first set of simpler models. The simulations indicate that V723 Cas central source achieved $L = 10^{38}$ $erg/s$ and $T_{eff} = 280,000$ $K$ in 2008, consistently with the X-ray observations of $2.8 \times 10^5 \leq T_{eff} \leq 3.8 \times 10^5$ $K$ and $5 \times 10^{36} \leq L \leq 2 \times 10^{38}$ $erg/s$ (Ness et al. 2008). These values are also needed to reproduce the order of magnitude of the permitted line fluxes. Using the initial assumptions of Al, Si and Ca abundances, we could not reproduce the forbidden lines fluxes or the lines flux ratios, as showed in figure 6. Therefore, we assumed that those abundances were underestimated (see discussion in next section), and we ran new models with higher values. The set of abundances that best fitted the observational fluxes are $log(N_{Al}/N_H) = -5.4$, $log(N_{Si}/N_H) = -4.7$ and $log(N_{Ca}/N_H) = -6.4$. Figure 7 displays the volume emissivity in the polar plane for each line for the best fitted RAINY3D model. It is possible to notice the different contrasts for each line, that are also observed in Keck-OSIRIS data. Compared to the previous model (with uniform ionizing source, fig. 4), the [Al IX] structures change significantly, indicating a better agreement to the observations. As these models include only a disk-shaped ionizing source, there is an equatorial region that is not ionized, represented by white gaps in figure 7.

Despite the reasonable fit, it is not reasonable to assume that an ionizing source composed only by an accretion disk achieves luminosity as high as $10^{38}$ $erg/s$. Assuming an upper limit of mass transfer from the secondary to the primary of $10^{-7}$ $M_{\odot}/year$, the luminosity of the disk would be around $10^{34}$ $erg/s$. The disk also scatters the light from the pseudo-photosphere, so the total luminosity of the disk can be higher than $10^{34}$ $erg/s$, but the presence of the spherical component is needed. Therefore, we tested a few new models with both spherical and disk components as ionizing sources, to estimate their contribution. Figure 8 shows the variation of the lines mean volume emissivity (in $erg/cm^3/s$) along the disk ($\theta = 90°$) and along the polar caps ($\theta = 0°$), as a function of the disk luminosity. Our results suggest that at least half of the ionizing luminosity must be emitted (and scattered) by the disk, and the other half must come from the spherical source. It is important to stress that this disk high luminosity may be possible due to the radiation repro-cessing effects that occur during supersoft phase (Popham & di Stefano 1996).

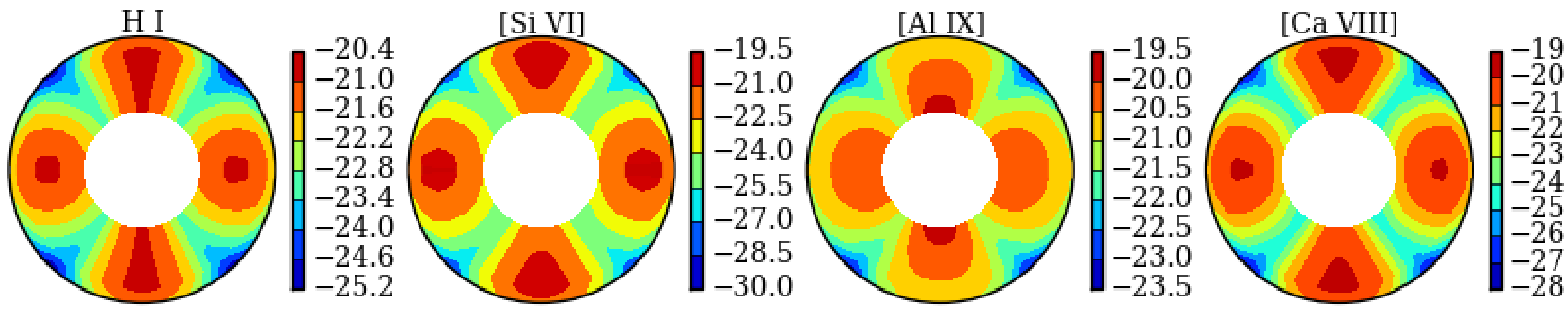

**Figure 4.** RAINY3D output local line emissivity for Brγ, [Al IX], [Si VI] and [Ca VIII] in the polar plane for models with a spherical central ionizing source. Emissivity units are $erg/cm^3/s$.

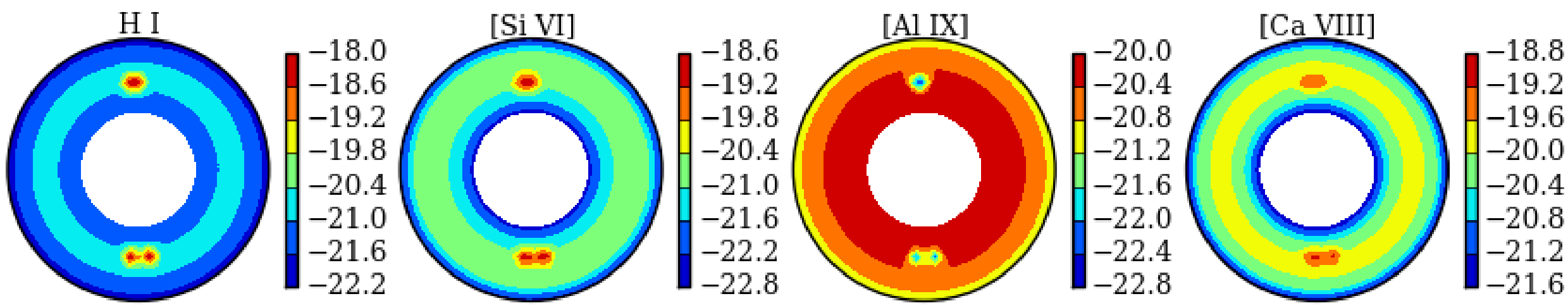

**Figure 5.** RAINY3D output of local emissivity for Brγ, [Al IX], [Si VI] and [Ca VIII] in the equatorial plane for models with a spherical central ionizing source.

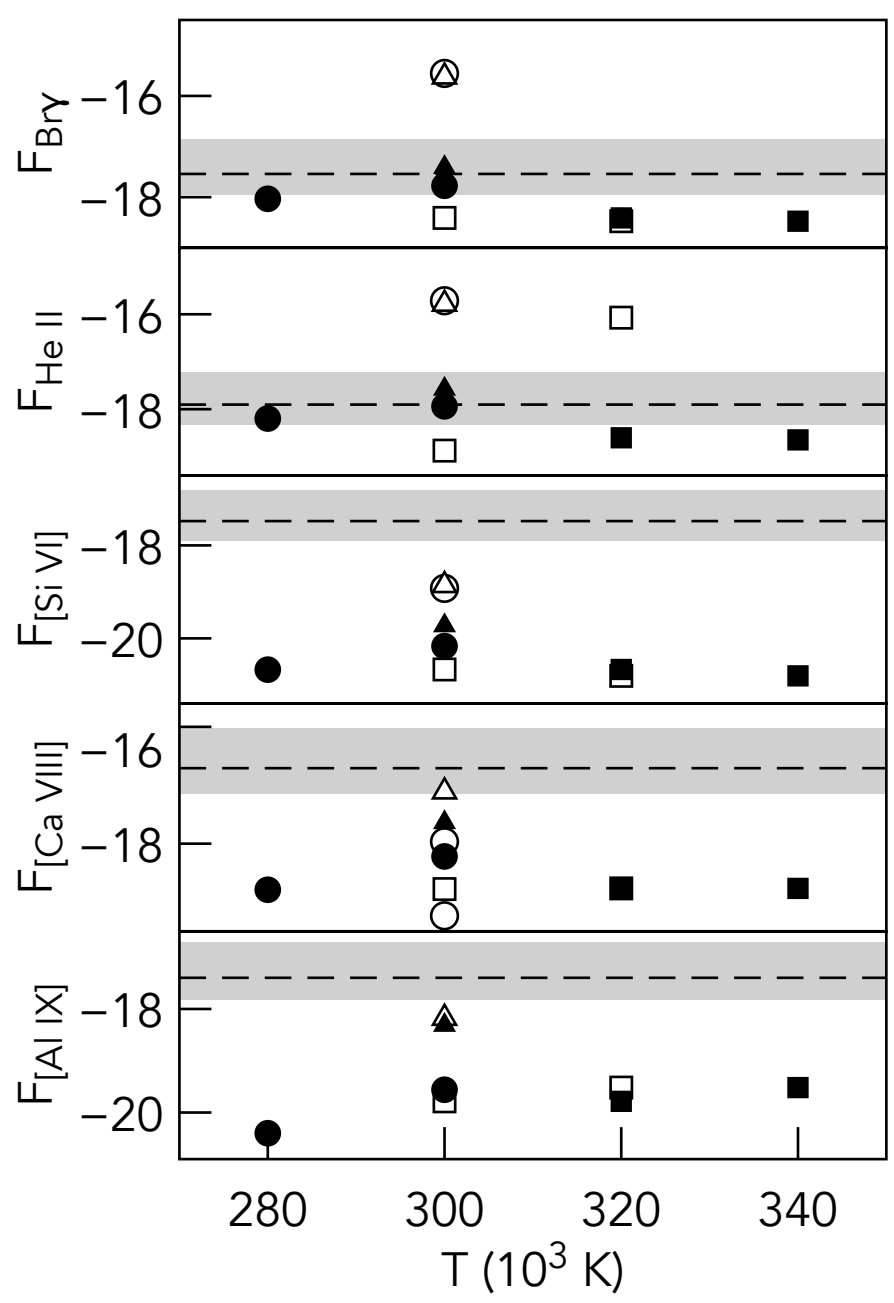

**Figure 6.** Results from models with 3D geometry and anisotropic ionizing radiation field, with $log(L) = 37.0$ (squares), $log(L) = 37.5$ (circles) and $log(L) = 38.0$ (triangles), where all L values are in $erg/s$. The white symbols are representing models with $M_H = 1.2 \times 10^{-5} M_\odot$ and the black ones are representing models with $M_H = 5.6 \times 10^{-6} M_\odot$.

The final models were made considering both disk-shaped and spherical ionizing sources, with same contribution to the total luminosity, in addition to the 3D shell geometry. The best fit was achieved with $M_{shell} = 1.1 \times 10^{-5} M_\odot$, $T_{c.s.} = 280,000\ K$ and $L_{c.s.} = 10^{38}\ erg/s$, $log(N_{Al}/N_H) = -5.4$, $log(N_{Ca}/N_H) = -6.4$ and $log(N_{Si}/N_H) = -4.7$. The resulting integrated line flux ratios from this model are shown in table 2 and the resulting projection of the line emissivity is displayed in figure 9. For the H line, it is important to stress that it is not possible to exactly compare the 2D observations to the projection of the modeled emissivity because the line self-absorption in the cloud is possibly relevant. Another problem of the H lines is the difficulty of decomposing the total emission flux into the central source and the shell contributions. Therefore, it is expected that the RAINY3D projections may present less similarities for the $Br\gamma$ line. For [Si VI] and [Ca VIII] lines, the ring and polar caps can be noticed in both observed images and model projections. For the [Al IX] line, the disk contribution to the shell radiation field is responsible for the bright emission from the polar caps, that also is present in both model projection and observed image. The ring and polar caps structures in our models are wider than the observed ones, due to the requirement of a smooth density gradient for fast convergence of gas pressure and cooling in the models.

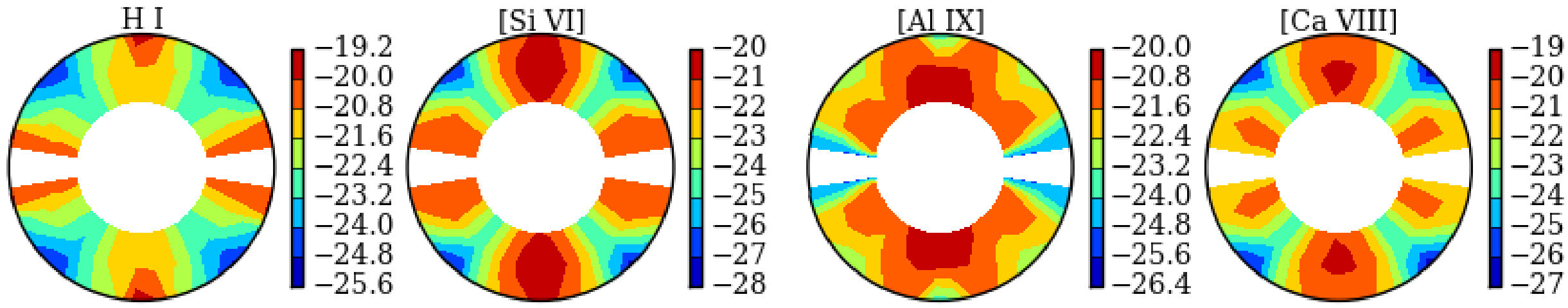

**Figure 7.** Model output of local emissivity for Brγ, [Al IX], [Si VI] and [Ca VIII] in the polar plane for models with a disk-shaped central ionizing source. The white gaps correspond to the non-ionized gas region, due to the disk source.

**Table 2.** Integrated fluxes ratios for Al, Si and Ca lines in the best fit 3D model

| | $F_{Si}/F_{Br\gamma}$ | $F_{Al}/F_{Br\gamma}$ | $F_{Ca}/F_{Br\gamma}$ | $F_{He}/F_{Br\gamma}$ | $F_{Al}/F_{Si}$ | $F_{Ca}/F_{Si}$ |
|---|---|---|---|---|---|---|
| Observed | 1.2 | 1.4 | 6.8 | 0.4 | 1.2 | 5.9 |
| Modeled | 1.0 | 1.3 | 6.2 | 0.5 | 1.3 | 6.1 |

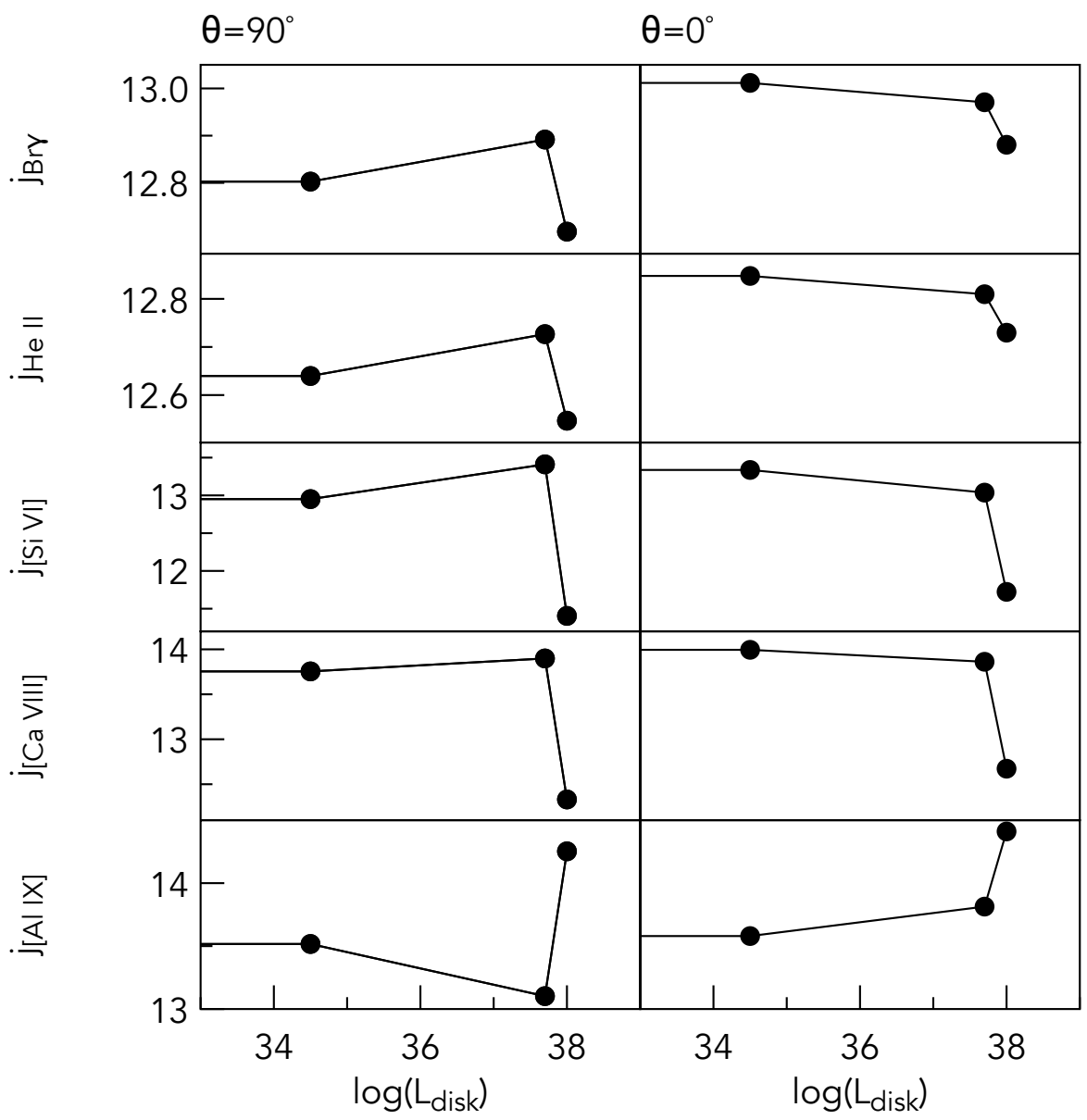

**Figure 8.** RAINY3D logarithm of mean volume emissivity (in $erg/cm^3/s$) for Brγ, [Al IX], [Si VI] and [Ca VIII] along polar and equatorial planes for different disk luminosity contribution.

## 4 DISCUSSION

### 4.1 3D geometry and anisotropic radiation field

Our models show the importance of considering a realistic density profile and a nonuniform ionizing radiation field (when indicated by the observations) for nova shells photoionization models. Depending on the geometry adopted, the line flux ratios can change significantly, especially the permitted to forbidden lines ratios.

The nonuniform ionizing field affects directly the central source parameters and consequently the lines fluxes. The presence of a disk-shaped ionizing field, for instance, leads to higher values of luminosity and/or effective temperature of the central source. When assuming this disk-shaped source, the ionizing radiation becomes concentrated in the poles, creating a cone with higher emissivity. Simultaneously, the contribution from the shell equatorial region becomes almost insignificant to the total emissivity.

In V723 Cas particular case, we checked if the polar structures observed in OSIRIS data could be reproduced only with the disk-shaped field, without the need of a mass concentration in the poles. We adopted a new 3D mass distribution formed by a spherical shell diffuse component and a clumpy equatorial ring component. Both structures had their masses increased compared to the former mass distribution, to ensure the total mass value would fall in the adopted mass range. The results did not show a significant polar structure, suggesting that both disk-shaped ionizing source and a polar density enhancement are needed to reproduce the observational features.

Our initial assumption of a disk ionizing source geometry could not be entirely correct because the luminosity of our models ($L = 10^{38}$ $erg/s$) was too high for an accretion disk. On the other hand, it is possible that accretion sets in while the nova burning is on (Starrfield et al. 2004). Therefore, at high mass transfer rates, the ionizing source is likely composed by an accretion disk and a spherical pseudo-photosphere. This is a reasonable assumption, since the X-ray data showed that hydrogen burning was still active in 2008. The proposed scenario is consistent with the hypothesis of V723 Cas being transitioning from a nova to a SSS phase (Ness et al. 2008) which lasted until 2014 (Ness et al. 2015).

### 4.2 Total emitted spectra

Our models provide information on the emission lines emissivity along an equatorial profile and along a polar profile, based on Cloudy atomic data. These files contain all the emission lines computed and can be used to check the line identification in the observed spectra. The physical condi-tions in the shell indicate that the line at 2.206 $\mu m$ identified as [Ti VII] (Rudy et al. 2002) is actually a [Fe XII] (2.206 $\mu m$) transition, and that an unidentified line at 2.2188 $\mu m$ (Rudy et al. 2002) is a [Fe IX] (2.218 $\mu m$) transition. The [Fe XII] ionization energy is 330.8 eV and the [Fe IX] one is 233.6 eV, which are similar to the ionization energies of [Al IX] (330.2 eV) and [Si VI] (205.3 eV). The Cloudy atomic database does not include the Ti coronal lines. On the other hand, according to NIST data (Kramida et al. 2015), there is a [Fe XII] line at 2.2178 $\mu m$, not a [Fe IX], and its transition probability is similar to the [Ca VIII] line present in OSIRIS spectra. Lyke & Campbell (2009) identified the 2.2188 $\mu m$ at 2.2190 $\mu m$, and they have pointed out that the emission from this line was similar to the [Al IX] emission, which lead us to suspect that the 2.2178 $\mu m$ line is, in fact, a [Fe XII] line.

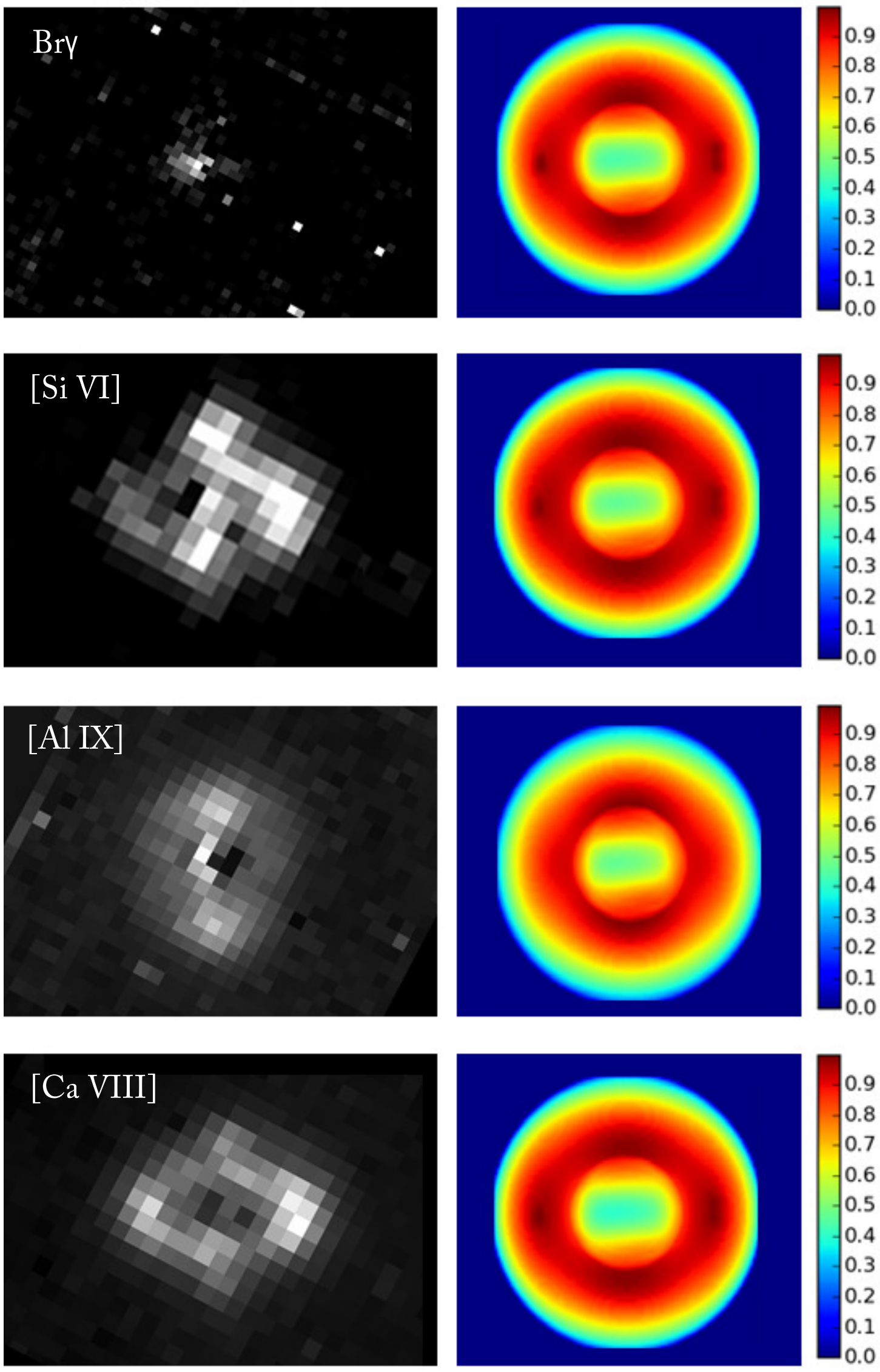

**Figure 9.** Best-fit model 2D surface brightness distributions (right) compared to Keck-OSIRIS observed data (left).

### 4.3 Si, Ca and Al abundances

We could not reproduce the observed spectra using the first assumption of Si, Ca and Al abundances. We varied the geometry, the mass distribution, the total mass value and the ionizing field, but all indicate that the Si, Ca and Al abundances must be higher than the ones found by Evans et al. (2003). An increase in mass, as well as an increase in temperature or luminosity of the central source, leads to higher permitted lines fluxes. Changes in density contrast for the geometry structures are also not enough to achieve the observed fluxes. The errors in distance and E(B-V) are also insufficient to be the cause of the observed to modeled fluxes deviation. The only way we could fit the observed fluxes was increasing the Si, Ca and Al abundances by ~ 1 *dex*. This is, actually, a reasonable assumption, considering that his abundances values are uncomfortably lower than the solar ones. These elements come from the secondary transferred gas, and it is reasonable that they are comparable to or even higher than the solar abundances.

### 4.4 V723 Cas as a neon nova

Iijima (2006) derived heavy metal abundances for V723 Cas from the optical nebular spectra obtained between 1996 and 1999. He found $log(N_{Ne}/N_H) = -2.5$, which is roughly 37 times the solar value, adopting $log(N_{Ne}/N_H)_\odot = -4.07$ (Asplund et al. 2009). This neon overabundance is typical of neon novae, as found in V5583 Sgr, V2491 Cyg, V1065 Cen, V382 Vel, V1974 Cyg, V1494 Aql, V4160 Sgr, V838 Her, V351 Pup, QU Vul, nova LMC 1990a and V693 CrA. IR spectra taken between 1995 and 1998 by Evans et al. (2003) did not show any neon lines, but the critical densities for Ne lines in IR are are almost 100 times lower than the critical density for Ne UV lines detected by Iijima (Sparke & Gallagher 2007). On the other hand, V723 Cas is a slow nova, and models of its light curve suggest a CO white dwarf of $\sim 0.5M_\odot$ (Hachisu & Kato 2015), while an O-Ne-Mg white dwarf is expected to have at least $1.2M_\odot$. The only way an O-Ne-Mg white dwarf would present a lower mass would be a mass loss after its formation. Kepler et al. (2016) found a white dwarf with an oxygen atmosphere with only 0.56 $M_\odot$ and discuss how this object would have lost its H+He shell. They propose a late violent thermal pulse in post-AGB phase, a C-shell flash and effects of a binary evolution as possible scenarios. In V723 Cas, we may be looking at a similar rare object - an O-Ne-Mg white dwarf that has been eroded during its binary evolution.

Our models suggest that V723 Cas luminosity in 2008 was $\sim 10^{38}$ $erg/s$, which is the Eddington luminosity for a $1M_\odot$ star, indicating the presence of a more massive white dwarf. On the other hand, the presence of metals in the gas and the non-spherical accretion due to the disc influence this limit, making it possible for a lower mass white dwarf to present this luminosity.

Is it reasonable to assume that if V723 Cas is a neon nova, it should have presented a strong neon emission line in our K band OSIRIS spectra? According to Kramida et al. (2015), there are only a few [Ne VIII] (239 $eV$) transitions in the K band. They do not have high Einstein coefficients and we could not find any neon nova K band spectrum with these emission lines.

One argument that could explain the slow light curve of V723 Cas in the presence of high mass white dwarf would be a low CNO abundance. As catalysts in the TNR reaction cycle, these elements limit the energy production at the nuclear burning peak and a deficiency of CNO would imply a less efficient burning. Opposing this scenario, the faint or absent H and He lines and respectively low model abundances in V723 Cas are intriguing. If most of the ejected material come from a normal secondary star, one may expect a higher abundance of H and He. The lack of observed H and He may suggest that, actually, there is a significant dredge-up of material from the white dwarf. Thus, the observed shell would be composed by the donor H-rich material diluted in the white dwarf CO, or O-Ne-Mg.

## 5 CONCLUSIONS

Our RAINY3D photoionization models based on Keck-OSIRIS data indicate that V723 Cas had an ionizing source with both spherical and disk components, with a substantial contribution the disk-like source being essential to match the 2d observations. A high temperature $280,000K$ central source is found for this nova and the total luminosity approaches the Eddington limit $L = 10^{38}$ $erg/s$. The models also suggest higher values for Al, Ca and Si abundances than previous literature findings, being $log(N_{Al}/N_H) = -5.4$, $log(N_{Ca}/N_H) = -6.4$ and $log(N_{Si}/N_H) = -4.7$. We estimated the total ejected shell mass in $M_{shell} = 1.1 \times 10^{-5}$ $M_\odot$.

The evidence that V723 Cas is a neon nova is sound, mainly due to its well defined neon overabundance, despite the fact that it hasn't been classified as one. Previous white dwarf mass loss episode(s) may conciliate the light-curve indication of a 0.6 $M_\odot$ white dwarf and the measured Ne overabundance.

## ACKNOWLEDGEMENTS

We thank FAPESP for the support under grant 2014/10326-3 and CNPq funding under grant #305657. We would also like to thank V. A. R. M. Ribeiro for his ideas to this project.

## APPENDIX A: RESULTS FROM RAINY3D MODELS

**Table A1.** Line flux ratios for the first grid of models with $log(L) = 36.5$

| $m_{shell}$ ($M_{\odot}$) | Temperature ($\times 10^3 K$) | $F_{Si}/F_{Br\gamma}$ | $F_{Al}/F_{Br\gamma}$ | $F_{Ca}/F_{Br\gamma}$ | $F_{He}/F_{Br\gamma}$ | $F_{Al}/F_{Si}$ | $F_{Ca}/F_{Si}$ |
|---|---|---|---|---|---|---|---|
| | | 1.16E+00 | 1.39E+00 | 6.80E+00 | 4.35E-01 | 1.20E+00 | 5.88E+00 |
| 6.00E-05 | 280 | 8.61E-03 | 2.57E-02 | 7.01E-01 | 6.44E-01 | 2.99E+00 | 8.14E+01 |
| 6.00E-05 | 300 | 5.24E-03 | 6.97E-02 | 6.85E-01 | 6.24E-01 | 1.33E+01 | 1.31E+02 |
| 6.00E-05 | 320 | 3.18E-03 | 1.43E-01 | 6.13E-01 | 6.07E-01 | 4.51E+01 | 1.93E+02 |
| 6.00E-05 | 340 | 1.96E-03 | 2.41E-01 | 5.18E-01 | 5.84E-01 | 1.23E+02 | 2.64E+02 |
| 6.00E-05 | 360 | 1.25E-03 | 3.55E-01 | 4.22E-01 | 5.57E-01 | 2.85E+02 | 3.39E+02 |
| 6.00E-05 | 380 | 8.20E-04 | 4.71E-01 | 3.35E-01 | 5.33E-01 | 5.75E+02 | 4.08E+02 |
| 8.49E-05 | 280 | 7.91E-03 | 1.39E-02 | 5.64E-01 | 6.52E-01 | 1.75E+00 | 7.13E+01 |
| 8.49E-05 | 300 | 7.96E-03 | 3.14E-02 | 6.31E-01 | 6.40E-01 | 3.95E+00 | 7.92E+01 |
| 8.49E-05 | 320 | 5.28E-03 | 7.48E-02 | 6.46E-01 | 6.14E-01 | 1.42E+01 | 1.22E+02 |
| 8.49E-05 | 340 | 3.41E-03 | 1.37E-01 | 5.97E-01 | 5.87E-01 | 4.02E+01 | 1.75E+02 |
| 8.49E-05 | 360 | 2.27E-03 | 2.16E-01 | 5.32E-01 | 5.61E-01 | 9.54E+01 | 2.34E+02 |
| 8.49E-05 | 380 | 1.54E-03 | 3.04E-01 | 4.55E-01 | 5.37E-01 | 1.97E+02 | 2.96E+02 |
| 1.20E-04 | 280 | 3.33E-04 | 8.46E-06 | 1.27E-02 | 6.87E-01 | 2.54E-02 | 3.80E+01 |
| 1.20E-04 | 300 | 7.87E-03 | 1.89E-02 | 5.47E-01 | 6.42E-01 | 2.40E+00 | 6.95E+01 |
| 1.20E-04 | 320 | 5.23E-03 | 2.54E-02 | 4.01E-01 | 6.44E-01 | 4.85E+00 | 7.66E+01 |
| 1.20E-04 | 340 | 5.80E-03 | 6.99E-02 | 5.90E-01 | 6.06E-01 | 1.21E+01 | 1.02E+02 |
| 1.20E-04 | 360 | 4.02E-03 | 1.20E-01 | 5.69E-01 | 5.78E-01 | 2.99E+01 | 1.41E+02 |
| 1.20E-04 | 380 | 2.84E-03 | 1.80E-01 | 5.26E-01 | 5.51E-01 | 6.35E+01 | 1.85E+02 |

**Table A2.** Line flux ratios for the first grid of models with $log(L) = 37.0$

| $m_{shell}$ ($M_{\odot}$) | Temperature ($\times 10^3 K$) | $F_{Si}/F_{Br\gamma}$ | $F_{Al}/F_{Br\gamma}$ | $F_{Ca}/F_{Br\gamma}$ | $F_{He}/F_{Br\gamma}$ | $F_{Al}/F_{Si}$ | $F_{Ca}/F_{Si}$ |
|---|---|---|---|---|---|---|---|
| | | 1.16E+00 | 1.39E+00 | 6.80E+00 | 4.35E-01 | 1.20E+00 | 5.88E+00 |
| 6.00E-05 | 280 | 1.75E-03 | 2.15E-01 | 4.29E-01 | 6.49E-01 | 1.23E+02 | 2.45E+02 |
| 6.00E-05 | 300 | 8.80E-04 | 4.16E-01 | 2.82E-01 | 6.42E-01 | 4.73E+02 | 3.20E+02 |
| 6.00E-05 | 320 | 4.59E-04 | 6.59E-01 | 1.76E-01 | 6.26E-01 | 1.44E+03 | 3.83E+02 |
| 6.00E-05 | 340 | 2.51E-04 | 9.06E-01 | 1.07E-01 | 6.10E-01 | 3.60E+03 | 4.27E+02 |
| 6.00E-05 | 360 | 1.44E-04 | 1.12E+00 | 6.51E-02 | 5.93E-01 | 7.76E+03 | 4.52E+02 |
| 6.00E-05 | 380 | 8.46E-05 | 1.26E+00 | 3.93E-02 | 5.69E-01 | 1.50E+04 | 4.65E+02 |
| 8.49E-05 | 280 | 3.01E-03 | 1.27E-01 | 5.86E-01 | 6.45E-01 | 4.23E+01 | 1.95E+02 |
| 8.49E-05 | 300 | 1.57E-03 | 2.65E-01 | 4.32E-01 | 6.31E-01 | 1.69E+02 | 2.75E+02 |
| 8.49E-05 | 320 | 8.55E-04 | 4.51E-01 | 3.03E-01 | 6.16E-01 | 5.28E+02 | 3.54E+02 |
| 8.49E-05 | 340 | 4.85E-04 | 6.54E-01 | 2.04E-01 | 6.00E-01 | 1.35E+03 | 4.22E+02 |
| 8.49E-05 | 360 | 2.87E-04 | 8.51E-01 | 1.34E-01 | 5.80E-01 | 2.97E+03 | 4.68E+02 |
| 8.49E-05 | 380 | 1.75E-04 | 1.01E+00 | 8.69E-02 | 5.59E-01 | 5.75E+03 | 4.96E+02 |
| 1.20E-04 | 280 | 4.92E-03 | 6.89E-02 | 6.97E-01 | 6.42E-01 | 1.40E+01 | 1.42E+02 |
| 1.20E-04 | 300 | 2.72E-03 | 1.58E-01 | 5.77E-01 | 6.27E-01 | 5.80E+01 | 2.12E+02 |
| 1.20E-04 | 320 | 1.53E-03 | 2.87E-01 | 4.50E-01 | 6.07E-01 | 1.87E+02 | 2.93E+02 |
| 1.20E-04 | 340 | 9.02E-04 | 4.44E-01 | 3.36E-01 | 5.89E-01 | 4.92E+02 | 3.73E+02 |
| 1.20E-04 | 360 | 5.54E-04 | 6.10E-01 | 2.44E-01 | 5.71E-01 | 1.10E+03 | 4.42E+02 |
| 1.20E-04 | 380 | 3.49E-04 | 7.60E-01 | 1.73E-01 | 5.49E-01 | 2.17E+03 | 4.96E+02 |

**Table A3.** Line flux ratios for first grid of models with $log(L) = 37.5$

| $m_{shell}$ ($M_{\odot}$) | Temperature ($\times 10^3 K$) | $F_{Si}/F_{Br\gamma}$ | $F_{Al}/F_{Br\gamma}$ | $F_{Ca}/F_{Br\gamma}$ | $F_{He}/F_{Br\gamma}$ | $F_{Al}/F_{Si}$ | $F_{Ca}/F_{Si}$ |
|---|---|---|---|---|---|---|---|
| | | 1.16E+00 | 1.39E+00 | 6.80E+00 | 4.35E-01 | 1.20E+00 | 5.88E+00 |
| 6.00E-005 | 280 | 2.35E-04 | 7.95E-01 | 6.76E-02 | 6.67E-01 | 3.38E+03 | 2.87E+02 |
| 6.00E-005 | 300 | 1.02E-04 | 1.23E+00 | 3.07E-02 | 6.64E-01 | 1.21E+04 | 3.02E+02 |
| 6.00E-005 | 320 | 4.60E-05 | 1.57E+00 | 1.44E-02 | 6.50E-01 | 3.40E+04 | 3.13E+02 |
| 6.00E-005 | 340 | 2.17E-05 | 1.80E+00 | 7.16E-03 | 6.37E-01 | 8.32E+04 | 3.30E+02 |
| 6.00E-005 | 360 | 1.05E-05 | 1.89E+00 | 3.75E-03 | 6.21E-01 | 1.80E+05 | 3.58E+02 |
| 6.00E-005 | 380 | 5.13E-06 | 1.85E+00 | 2.06E-03 | 5.98E-01 | 3.60E+05 | 4.02E+02 |
| 8.49E-005 | 280 | 4.49E-04 | 5.74E-01 | 1.32E-01 | 6.64E-01 | 1.28E+03 | 2.94E+02 |
| 8.49E-005 | 300 | 2.02E-04 | 9.37E-01 | 6.56E-02 | 6.56E-01 | 4.64E+03 | 3.25E+02 |
| 8.49E-005 | 320 | 9.63E-05 | 1.29E+00 | 3.29E-02 | 6.45E-01 | 1.33E+04 | 3.42E+02 |
| 8.49E-005 | 340 | 4.78E-05 | 1.56E+00 | 1.70E-02 | 6.34E-01 | 3.25E+04 | 3.56E+02 |
| 8.49E-005 | 360 | 2.45E-05 | 1.71E+00 | 9.11E-03 | 6.13E-01 | 6.96E+04 | 3.71E+02 |
| 8.49E-005 | 380 | 1.29E-05 | 1.74E+00 | 5.11E-03 | 5.89E-01 | 1.35E+05 | 3.97E+02 |
| 1.20E-004 | 280 | 8.26E-04 | 3.89E-01 | 2.37E-01 | 6.58E-01 | 4.70E+02 | 2.87E+02 |
| 1.20E-004 | 300 | 3.89E-04 | 6.80E-01 | 1.31E-01 | 6.50E-01 | 1.75E+03 | 3.37E+02 |
| 1.20E-004 | 320 | 1.93E-04 | 9.96E-01 | 7.16E-02 | 6.40E-01 | 5.16E+03 | 3.70E+02 |
| 1.20E-004 | 340 | 1.01E-04 | 1.27E+00 | 3.92E-02 | 6.24E-01 | 1.26E+04 | 3.89E+02 |
| 1.20E-004 | 360 | 5.43E-05 | 1.47E+00 | 2.17E-02 | 6.04E-01 | 2.70E+04 | 4.01E+02 |
| 1.20E-004 | 380 | 3.01E-05 | 1.56E+00 | 1.25E-02 | 5.81E-01 | 5.20E+04 | 4.14E+02 |

**Table A4.** Line flux ratios for the second grid of models

| Temperature ($\times 10^3 K$) | $F_{Si}/F_{Br\gamma}$ | $F_{Al}/F_{Br\gamma}$ | $F_{Ca}/F_{Br\gamma}$ | $F_{He}/F_{Br\gamma}$ | $F_{Al}/F_{Si}$ | $F_{Ca}/F_{Si}$ |
|---|---|---|---|---|---|---|
| | 1.16E+00 | 1.39E+00 | 6.80E+00 | 4.35E-01 | 1.20E+00 | 5.88E+00 |
| 260 | 4.88E-03 | 1.47E-02 | 3.40E-01 | 6.85E-01 | 3.02E+00 | 6.96E+01 |
| 300 | 1.44E-03 | 4.75E-03 | 6.99E-02 | 6.89E-01 | 3.30E+00 | 4.86E+01 |
| 320 | 1.12E-03 | 5.64E-03 | 5.50E-02 | 6.81E-01 | 5.04E+00 | 4.91E+01 |
| 360 | 2.63E-03 | 1.91E-01 | 3.46E+00 | 6.07E-01 | 7.27E+01 | 1.32E+03 |

**Table A5.** Line flux ratios for the third grid of models

| $log(L)$ ($erg/s$) | Temperature ($\times 10^3 K$) | $F_{Si}/F_{Br\gamma}$ | $F_{Al}/F_{Br\gamma}$ | $F_{Ca}/F_{Br\gamma}$ | $F_{He}/F_{Br\gamma}$ | $F_{Al}/F_{Si}$ | $F_{Ca}/F_{Si}$ |
|---|---|---|---|---|---|---|---|
| | | 1.16E+00 | 1.39E+00 | 6.80E+00 | 4.35E-01 | 1.20E+00 | 5.88E+00 |
| 37.0 | 300 | 4.28E-04 | 2.10E-04 | 3.88E-03 | 6.93E-01 | 4.90E-01 | 9.06E+00 |
| 37.0 | 320 | 6.89E-04 | 8.93E-05 | 1.37E-03 | 6.78E-01 | 1.30E-01 | 1.99E+00 |
| 37.0 | 340 | 4.66E-03 | 9.05E-02 | 5.12E-01 | 6.73E-01 | 1.94E+01 | 1.10E+02 |
| 37.5 | 280 | 2.29E-03 | 4.29E-03 | 1.75E-01 | 6.84E-01 | 1.88E+00 | 7.66E+01 |
| 37.5 | 300 | 4.06E-03 | 1.64E-02 | 3.60E-01 | 6.84E-01 | 4.04E+00 | 8.88E+01 |